\documentclass[review]{elsarticle}

\usepackage{lineno,hyperref}
\usepackage{amsmath}
\usepackage{float}
\usepackage{subfigure}
\usepackage{xcolor}
\usepackage{bm}
\usepackage[normalem]{ulem}
\usepackage{cancel}
\newcommand{\fn}[1]{\left( #1 \right)}

\newcommand{\abs}[1]{\left| #1 \right|}
\newcommand{\epf}{\fn{1-\phi}}

\newcommand{\bld}[1]{{\bf #1}}

\modulolinenumbers[5]

\journal{International Journal of Multiphase Flow}

\usepackage{numcompress}\biboptions{numbers,sort&compress} 

\begin{document}

\begin{frontmatter}

\title{Fully resolved simulation of dense suspensions of freely evolving buoyant particles using an improved immersed boundary method}
\tnotetext[mytitlenote]{This is a post-peer-review, pre-copyedit version of an article published in International Journal of Multiphase Flow. The final authenticated version is available online at: \href{https://doi.org/10.1016/j.ijmultiphaseflow.2020.103396} {https://doi.org/10.1016/j.ijmultiphaseflow.2020.103396}.\\ \copyright 2020. This manuscript version is made available under the CC-BY-NC-ND 4.0 license \href{http://creativecommons.org/licenses/by-nc-nd/4.0/}{http://creativecommons.org/licenses/by-nc-nd/4.0/}.}


\author[firstaddress,secondaddress]{Vahid Tavanashad}

\author[firstaddress,secondaddress]{Shankar Subramaniam\corref{corres_author}}
\cortext[corres_author]{Corresponding author}
\ead{shankar@iastate.edu}

\address[firstaddress]{Department of Mechanical Engineering, Iowa State University, Ames, IA 50011, USA}
\address[secondaddress]{Center for Multiphase Flow Research \& Education (CoMFRE), Iowa State University, Ames, IA 50011, USA}

\begin{abstract}
Fully resolved simulation of flows with buoyant particles is a challenging problem since buoyant particles are lighter than the surrounding fluid, and as a result, the two phases are strongly coupled together.
In this work, the virtual force stabilization technique introduced by Schwarz et al. [Schwarz, S., Kempe, T., \& Fr{\"o}hlich, J. (2015). A temporal discretization scheme to compute the motion of light particles in viscous flows by an immersed boundary method. J. Comput. Phys., 281, 591--613] is extended to simulate buoyant particle suspensions with high volume fractions (up to $40 \%$).
It is concluded that the dimensionless numerical model constant $C_v$ in the virtual force technique should increase with volume fraction.
The behavior of a single rising particle, two in-line rising particles, and buoyant particle suspensions are studied.
In each case, results are compared with experimental works on bubbly flows to highlight the differences and similarities between buoyant particles and bubbles.
Finally, the drag coefficient is extracted from simulations of buoyant particle suspensions at different volume fractions, and based on that a drag correlation is presented.
\end{abstract}

\begin{keyword}
contaminated bubble \sep buoyant particle \sep immersed boundary method \sep fully-resolved direct numerical simulation \sep drag coefficient \sep drag law
\end{keyword}

\end{frontmatter}

\section{Introduction}
Free fall and rise of solid particles in a fluid is a type of dispersed multiphase flow.
This type of flow is of interest in numerous fields, including chemical, mechanical, and environmental engineering.
In particular, many types of microplastics, such as polyethylene and polypropylene, are considered buoyant in the oceans unless they are altered by biofilm growth \citep{wright_2013,driedger_2015}.
Besides, the simulation of buoyant particles can be used as an approximation for bubbles under certain conditions.
An important factor that affects the flow physics in bubbly flows is the purity of the surrounding liquid \citep{clift_book,magnaudet_2000,takagi_Matsumoto_2011}.
It is shown in many experimental studies \citep{sridhar_1995,belFdhila_1996, cartellier_2001,dijkhuizen_2010b,peters_Els_2012} that the rise velocity of a single bubble is significantly higher in clean systems, such as pure water, when compared to contaminated systems, such as tap water, especially in parameter ranges where the shape of the bubble is spherical or ellipsoidal.
Moreover, these studies show that small spherical bubbles in contaminated systems behave like solid spheres.
This means that the simulation of bubbles with the no-slip velocity boundary condition at the interface of two phases, instead of free-slip, is a good approximation of contaminated bubbly flows.

Fully-resolved direct numerical simulations (FR-DNS) of particle-laden flows are used as a tool for discovering flow physics as well as model development for macro-scale simulations such as Eulerian--Eulerian or Eulerian--Lagrangian approaches \citep{tenneti_arfm_2014}.
Therefore, it is necessary to develop numerical methods capable of simulating systems that span a wide range of particle-to-fluid density ratios from heavy to light particles.
Over time, different numerical methods have been developed and improved for FR-DNS of particle-laden flows such as immersed boundary \citep{uhlmann_2005,tenneti_pt_2010,kempe_2012,breugem_2012}, lattice Boltzmann \citep{hill_koch_2001a,hill_koch_2001b,vanderHoef_2005,yin_koch_2007,yin_koch_2008}, fictitious domain with Lagrange multipliers \citep{glowinski_1999,sharma_2005,yu_2006,apte_2009,apte_2013}, PHYSALIS \citep{zhang_prosperetti_2005,sierakowski_2016}, and body-fitted \citep{hu_2001,burton_eaton_2005,bagchi_2002a,bagchi_2002b} methods. 
In these methods, the motion of particles is determined by Newton's equations of motion.

When loose coupling schemes are used in these methods, the interface boundary conditions may not be satisfied accurately since these schemes only involve the solution of the fluid and the particle, once per time step.
This incompatibility of the kinematic and dynamic quantities at the interface may cause severe stability issues when the particle density $\rho_p$ is close to, or smaller than, the fluid density $\rho_f$ \citep{inamuro_2003,uhlmann_2005,yin_koch_2008,kempe_2012, breugem_2012,apte_2013,yang_2014,maxey_arfm_2017}.
This numerical instability is known as artificial added-mass effects of the fluid on the particle, and is common in fluid--structure interaction problems \citep{causin_2005,forster_2007,borazjani_2008,sotiropoulos_2014}.
Although the use of strong coupling methods (implicit coupling schemes) solves the instability problem at the expense of computational time \citep{hu_2001,fernandez_2005,figueroa_2006}, numerous studies have attempted to stabilize explicit coupling schemes to overcome the problem and benefit from the simplicity of explicit methods \citep{tallec_2001,burman_2009,guidoboni_2009,apte_2013}.

In the present study, we develop a stabilized explicit coupling scheme for FR-DNS using the immersed boundary method (IBM) that is capable of simulating buoyant rigid particles.
Several studies have shown that the IBM becomes unstable for low particle-to-fluid density ratios when explicit methods are used. \citep{uhlmann_2005,kempe_2012,breugem_2012,yang_2014,maxey_arfm_2017}.
For instance, the IBM developed by \citet{uhlmann_2005} becomes unstable when the density ratio $\rho_p/\rho_f$ is smaller than $1.2$.
To extend the stability range, \citet{kempe_2012} succeeded in performing simulations with density ratio as low as $0.3$ by numerical evaluation of volume integrals in the equation of particle motion, instead of using the rigid-body motion assumption (assuming that the fictitious fluid motion inside the particle is equal to rigid-body motion irrespective of the actual type of motion inside the volume) made by Uhlmann.
\citet{yang_2014} utilized the rigid-body assumption but used a 4th-order predictor-corrector scheme to solve the equation of motion for particles and achieved stable solution for $\rho_p/\rho_f > 0.29$.
Our implementation of IBM \citep{garg_book,tenneti_pt_2010} which is called the particle-resolved uncontaminated-fluid reconcilable immersed boundary method (PUReIBM)  directly calculates the hydrodynamic force on particle surface from the stress field, and it is stable for $\rho_p/\rho_f > 0.07$ (see section \ref{sec:single_time}).

While the works of \citet{kempe_2012} and \citet{yang_2014} ``improve'' the stability limit, \citet{tschisgale_2017} have developed a non-iterative implicit IBM to ``remove'' the low-density ratio restriction and they successfully simulated a single rising particle with density ratio $0.001$.
Another promising work for removing the low-density ratio restriction is done by \citet{schwarz_Kempe_2015}.
In their work, a virtual force stabilization technique is introduced, which allows simulation of a single rising particle with density ratio as low as $0.001$.
In this technique, the governing equations are solved using an explicit method, and a virtual force is added to the equation of particle motion to stabilize the method.
This idea was originally used for a single rising particle and later was used for very dilute systems of contaminated bubble swarms with volume fraction up to $2.14 \%$ \citep{santarelli_2015}.
The focus of the work by \citet{schwarz_Kempe_2015} is on numerical accuracy.
They tried to design a numerical scheme with the same accuracy as their original method without the virtual force.
The virtual force method has a dimensionless numerical model constant $C_v$, which represents the magnitude of the dimensionless virtual force in the particle equation of motion, and \citet{schwarz_Kempe_2015} claimed that this method works for any $C_v$ greater than zero.

In this paper, we show that there is a lower positive limit for $C_v$, which depends on the density ratio and the added mass coefficient.
This condition is essential in the case of buoyant particle suspensions compared to a single particle because the added mass coefficient is affected by the volume fraction.
Therefore, it is important to choose a $C_v$ that results in stable solutions for buoyant particle suspensions.
The ultimate goal of this work is to develop an explicit time-stepping method that simulates buoyant particles for arbitrary density ratio and a range of volume fractions.
In this regard, the results of a single rising particle are presented for validation. 
Then, the behavior of two in-line rising particles is studied.
Finally, the drag coefficient is extracted from simulations of buoyant particle suspensions at different volume fractions, as an illustrative example to show the applicability of this method in practice.
In each part, results are compared with experimental works on bubbly flows to highlight the differences and similarities between buoyant particles and bubbles.
Based on the data obtained from simulations of buoyant particle suspensions, we have also proposed a correlation for the drag coefficient of particle suspension to show the application of our modified FR-DNS solver for model development.

Although in this study we use IBM to solve the governing equations, the virtual force stabilization technique is not restricted to IBM and could be used in any flow solver in which the motion of bubbles or buoyant particles is determined by Newton's equations of motion.
For instance, \citet{xia_2019} used this technique to modify the direct-forcing fictitious domain method for particle-laden flows of arbitrary density ratio and dilute systems with volume fraction up to $0.84 \%$.

The rest of the paper is organized as follows: In section \ref{sec:instability}, the underlying reason for instability in the simulations of buoyant particles is explained. 
Then the virtual force stabilization technique is introduced, and a detailed explanation is provided as to why and how it works.
In section \ref{sec:method}, the IBM used in this work is introduced, and modifications required of the original method to incorporate the virtual force are explained.
In section \ref{sec:results}, the new method is validated, and simulation results for dense buoyant particle suspensions are presented.
Finally, conclusions are drawn in section \ref{sec:conclusion}.

\section{Numerical instability in simulations of strongly coupled particle-fluid flows}
\label{sec:instability}
Although the final goal of this study is to perform FR-DNS, for simplicity, the instability problem that occurs in low-density-ratio simulations is first explained through point particle equations in sections \ref{sec:point_particle} and \ref{sec:virtual_force}.
The virtual force stabilization approach is easier to explain in the point-particle context because the added mass force appears explicitly, whereas in FR-DNS it is a part of the integral of fluid stress at the particle surface.
The extension to FR-DNS is discussed in section \ref{sec:resolved}.

\subsection{Explanation of the stability problem in point-particle simulations}
\label{sec:point_particle}
In the point-particle approach, the interaction between particles and the surrounding flow is modeled.
In this case, the equation of motion for particles is described using the Basset-Boussinesq-Oseen (BBO) equation,
\begin{equation}
	\label{eq:point_particle}
	m_p\frac{d\mathbf{V}}{dt} = \mathbf{F}_D + \mathbf{F}_L + \mathbf{F}_{AM} + \mathbf{F}_{BH} + \mathbf{F}_B ,
\end{equation}
where $\mathbf{V}$ is the particle velocity, $m_p$ is the particle mass, $\mathbf{F}_D$ denotes the drag force, $\mathbf{F}_L$ the lift force, $\mathbf{F}_B$ the body forces, $\mathbf{F}_{AM}$ the added mass force and $\mathbf{F}_{BH}$ the Basset history force.

In this equation, the added mass force is defined as:
\begin{equation}
	\label{eq:added_mass}
	\mathbf{F}_{AM} = m_{am} \left(\frac{D\mathbf{u}}{Dt} - \frac{d\mathbf{V}}{dt}\right),
\end{equation}
where $\mathbf{u}$ is the fluid velocity at the particle location and $m_{am}$ is the added mass which is usually defined by a dimensionless coefficient $C_{am}$ as $m_{am} = C_{am} \rho_f V_p$, where $\rho_f$ is density of fluid and $V_p$ is volume of particle.
Added mass represents the inertia added to a particle as it accelerates (or decelerates) and moves (or deflects) a portion of its surrounding fluid.

\citet{hu_2001} have shown that Eq. (\ref{eq:point_particle}) is unstable when it is solved with an explicit time integration method, and the added mass exceeds the particle mass.
They assume that at the early stages of motion, only body forces and added mass force are important.
Furthermore, they assume that at the early stages, the fluid acceleration is much smaller than the particle acceleration.
Under these assumptions, Eq. (\ref{eq:point_particle}) simplifies to:
\begin{equation}
	\label{eq:Hu_eq1}
	m_p\frac{d\mathbf{V}}{dt} = \mathbf{F}_B - m_{am}\frac{d\mathbf{V}}{dt}.
\end{equation}
Note that $\mathbf{F}_B = \left(\rho_p - \rho_f\right) V_p \mathbf{g}$ is a constant driving force ($\rho_p$ is particle density and $\mathbf{g}$ is the acceleration due to the gravity).
Starting from an initial condition and solving this equation explicitly for the next time step, it can be shown that acceleration at time step $n$ is related to the initial acceleration through the following equation \citep{hu_2001},
\begin{equation}
	\label{eq:Hu_eq2}
	\frac{d\mathbf{V}}{dt}\left(t_n\right) = \frac{\displaystyle{1-\left(-\frac{m_{am}}{m_p}\right)^n}}{m_p+m_{am}}\mathbf{F}_B - \left(-\frac{m_{am}}{m_p}\right)^n\frac{d\mathbf{V}}{dt}\left(t_0\right).
\end{equation}
Equation (\ref{eq:Hu_eq2}) shows that the particle velocity oscillates with increasingly large amplitude when the added mass is larger than the mass of particle.
Therefore, the stability condition for Eq. (\ref{eq:Hu_eq2}) is:
\begin{equation}
	\label{eq:Hu_condition}
	m_p > m_{am} \Rightarrow \rho_p/\rho_f > C_{am},
\end{equation}
which means density ratio should be greater than the added mass coefficient to have stable solution.
In the next subsection, we describe how to remove this instability condition.

\subsection{Solving the instability problem using the virtual force technique}
\label{sec:virtual_force}
\citet{schwarz_Kempe_2015} introduced the virtual force technique to stabilize the equation of motion for particles in the case of low-density ratio.
They defined the virtual force as:
\begin{equation}
	\label{eq:virtual_force}
	\mathbf{F}_v = C_v \rho_f V_p \frac{d\mathbf{V}}{dt},
\end{equation}
with $C_v$ an appropriately chosen coefficient.
Although the virtual force is defined similar to the added mass force, it is a purely mathematical term designed to stabilize the temporal integration and does not have any physical meaning.

\citet{schwarz_Kempe_2015} added the virtual force to a ``generic'' test case which excluded the added mass force and concluded that $C_v > 0$.
Here, we add the virtual force to Eq. (\ref{eq:Hu_eq1}) which includes the added mass force and show that there is a lower positive limit for $C_v$.
Starting with Eq. (\ref{eq:Hu_eq1}) and adding the virtual force to this equation results in:
\begin{equation}
	\label{eq:VF_concept1}
	m_p \frac{d\mathbf{V}}{dt} + \mathbf{F}_v = \mathbf{F}_B - m_{am} \frac{d\mathbf{V}}{dt} + \mathbf{F}_v.
\end{equation}
By defining $m_p^{eq} = m_p + C_v \rho_f V_p$ and $m_{am}^{eq} = m_{am} - C_v \rho_f V_p$, this equation is re-written as:
\begin{equation}
	\label{eq:VF_concept2}
	m_p^{eq} \frac{d\mathbf{V}}{dt} = \mathbf{F}_B - m_{am}^{eq} \frac{d\mathbf{V}}{dt}.
\end{equation}

It is clear that Eqs. (\ref{eq:Hu_eq1}) and (\ref{eq:VF_concept2}) are similar and we conclude that the latter equation is only stable if $ m_p^{eq} > m_{am}^{eq} $ which simplifies to:
\begin{equation}
	\label{eq:cv_condition}
	\frac{m_p^{eq}}{m_{am}^{eq}} > 1 \Rightarrow \frac{\rho_p + C_v \rho_f}{\left( C_{am} - C_v \right) \rho_f}  > 1 \Rightarrow C_v > \frac{C_{am}}{2} - \frac{\rho_p}{2 \rho_f}.
\end{equation}
This result shows that there is a lower limit for $C_v$ which we call $C_v^{min}$, and it depends on the added mass coefficient and the density ratio.

It is known that the geometry of a particle and the presence of other particles or bounding walls can affect the added mass coefficient \citep{simcik_2008}, so it is expected that the physics of the problem affects the stability condition on $C_v$ through the added mass coefficient, which is confirmed in section \ref{sec:swarm} where we simulate buoyant particle suspensions.

\subsection{Using the virtual force technique in fully resolved simulations}
\label{sec:resolved}
In FR-DNS, particles are fully resolved by the grid, and the flow field on the surface of each particle is captured by solving the Navier-Stokes equations.
The hydrodynamic force on each particle is calculated by integrating the pressure and viscous stress fields over the particle surface, and the resulting equation of motion for each particle is,
\begin{equation}
	\label{eq:particle_resolved}
	m_p\frac{d\mathbf{V}}{dt} = \mathbf{F}_h + \mathbf{F}_B,
\end{equation}
where $ \mathbf{F}_h = \displaystyle{\oint_{\Gamma_p}{\boldsymbol{\tau} \cdot \mathbf{n} ds}} $ is the hydrodynamic force, $\boldsymbol\tau = -\mathbf{I} p + \mu_f \left( \nabla \mathbf{u} + \nabla \mathbf{u}^T \right)$ is the hydrodynamic stress tensor, $\mathbf{I}$ the identity matrix, $\mu_f$ the fluid dynamic viscosity, $p$ is the pressure with the hydrostatic part being subtracted, and $\mathbf{n}$ is the normal vector at the surface of particle.

As mentioned in the Introduction, the solution of this equation also becomes problematic for low-density ratios as reported in the literature \citep{uhlmann_2005,kempe_2012,breugem_2012,yang_2014,maxey_arfm_2017}.
Similarly, our code, PUReIBM, also becomes unstable for low-density ratios, but only for $\rho_p/\rho_f < 0.07$. 
Therefore, it is proposed to apply the virtual force technique in PUReIBM to stabilize the fully resolved simulations for low-density ratios, which is the topic of section \ref{sec:method}.
Before explaining the detail of PUReIBM, we emphasize a few points about the virtual force technique:
\begin{enumerate}
\item In FR-DNS, the surface integral in Eq. (\ref{eq:particle_resolved}) is evaluated directly from the flow field yielding all forces introduced in Eq. (\ref{eq:point_particle}) acting on the particle.
\item Since Eq. (\ref{eq:particle_resolved}) inherently includes the added mass force, the virtual force technique can be used to stabilize this equation with a $C_v>C_v^{min}$.
\item  The value of $C_v^{min}$ is not necessarily exactly the same as found in Eq. (\ref{eq:cv_condition}) since Eqs. (\ref{eq:VF_concept1}) and (\ref{eq:particle_resolved}) are different.
However, $C_v^{min}$ still depends on the added mass coefficient and density ratio (more detail about this point is presented in section \ref{sec:results}).
\item The virtual force technique can be used to stabilize any explicit numerical method that solves Eq. (\ref{eq:particle_resolved}) coupled with fluid equations for low-density ratios and is not limited to IBM.
\end{enumerate}

\section{Numerical method for FR-DNS}
\label{sec:method}
The fully resolved simulation approach used in this work is based on the direct forcing immersed boundary method of \citet{jamalphd} which is further developed in \citep{garg_book,tenneti_pt_2010}, and is called the particle-resolved uncontaminated-fluid reconcilable immersed boundary method (PUReIBM). 
The PUReIBM methodology is explained in detail in other works \citep{tenneti_pt_2010,mehrabadi_jfm_2015,tenneti_jfm_2016} and has been extensively validated in different cases \cite{garg_book,tenneti_ijmf_2011,tenneti_jfm_2016}.
Here, the main features of this method are presented.

The governing equations of the fluid phase that are solved in PUReIBM are the continuity equation:
\begin{equation}
	\label{eq:cont_ibm}
	\boldsymbol{\nabla \cdot} \mathbf{u}=0,
\end{equation}
and the Navier-Stokes equations:
\begin{equation} 
	\label{eq:nse_ibm}
	\rho_f \frac{\partial \mathbf{u}} {\partial t} + \rho_f \boldsymbol{\nabla \cdot} {\left( {\mathbf{uu}} \right)} = - \nabla p + \mu_f \nabla^2 \mathbf{u} + \mathbf{f}_\text{IBM},
\end{equation}
which are solved on a uniform Cartesian grid points with the Crank-Nicolson scheme for the viscous terms and an Adams-Bashforth scheme for the convective terms.
The boundary conditions on the fluid velocity at the particle interface (no-slip and no-penetration) are imposed via the immersed boundary force term, $\mathbf{f}_\text{IBM}$.

The motion of each particle in PUReIBM is evolved by updating its position $\mathbf{X}$, translational velocity $\mathbf{V}$, and rotational velocity $\mathbf{\Omega}$ according to Newton's second law as:
\begin{eqnarray}
\frac{d \mathbf{X}}{dt}& = &\mathbf{V},
\label{eq:lag_pos}  \\
m_p \frac{d \mathbf{V}}{dt}& = &\mathbf{F}_B + \mathbf{F}_h + \sum_{\substack{{j=1}\\{j \neq i}}}^{N_p} \mathbf{F}_{c}^{(ij)},
\label{eq:lag_vel} \\
I_p \frac{d \mathbf{\Omega}}{dt}& = &\mathbf{T}_h + \sum_{\substack{{j=1}\\{j \neq i}}}^{N_p} \mathbf{T}_{c}^{(ij)},
\label{eq:rot_vel}
\end{eqnarray}
where $N_p$ is the number of particles, $\mathbf{F}_{c}^{(ij)}$ is the collisional force between the $i^{th}$ particle and $j^{th}$ particle, $I_p=1/10 \rho_p V_p d_p^2$ is the moment of inertia of particles, $d_p$ is the particle diameter, $\mathbf{T}_h$ is the hydrodynamic torque, and $\mathbf{T}_c$ is the collisional torque.

A soft-sphere collision \citep{cundall_dem} is used to model the particle-particle interactions. 
Particles are allowed to overlap during a collision, and the contact mechanics between the overlapping particles are modeled by a spring in the normal direction (elastic collisions).
The spring causes the colliding particles to rebound.
The particles considered in this study are assumed to be frictionless.
This implies that the tangential component of the contact force and $\mathbf{T}_c$ are zero.

In the next sections \ref{sec:pureibm_virtualforce} and \ref{sec:lubrication}, the modifications to the original PUReIBM to extend its capability to simulate buoyant particles are explained.

\subsection{Addition of virtual force and torque to PUReIBM}
\label{sec:pureibm_virtualforce}
Since Eq. (\ref{eq:lag_vel}) is solved explicitly for the particle acceleration in PUReIBM, some numerical instabilities arise in the code when the density ratio is small.
As discussed in the previous section, the virtual force (Eq. (\ref{eq:virtual_force})) should be added to both sides of Eq. (\ref{eq:lag_vel}) to stabilize the PUReIBM for low-density ratio simulations,
\begin{equation}
\label{eq:lag_vel_modified}
\left( m_p + C_v \rho_f V_p \right) \frac{d \mathbf{V}}{dt}  =  \mathbf{F}_B + \mathbf{F}_h + \sum_{\substack{{j=1}\\{j \neq i}}}^{N_p} \mathbf{F}_{c}^{(ij)} +  \mathbf{F}_v.
\end{equation}

Similar to the concept of virtual force, we can define a virtual torque $\mathbf{T}_v = 1/10 \; C^{T}_{v} \rho_f V_p d_p^2 \displaystyle{\frac{d \mathbf{\Omega}}{dt}}$ and add it to both sides of Eq. (\ref{eq:rot_vel}) to stabilize this equation,
\begin{equation}
\left( I_p + \frac{1}{10} \; C^{T}_{v} \; \rho_f V_p d_p^2 \right) \frac{d \mathbf{\Omega}}{dt} = \mathbf{T}_h + \sum_{\substack{{j=1}\\{j \neq i}}}^{N_p} \mathbf{T}_{c}^{(ij)} + \mathbf{T}_v.
\label{eq:rot_vel_modified}
\end{equation}
In the definition of virtual torque, $C^{T}_{v}$ is the virtual torque coefficient and it is considered to be equal to $C_v$ in this paper.

A combination of the Adams-Bashforth predictor-corrector scheme and the trapezoidal rule (see \ref{sec:appA}) is used to calculate $\mathbf{F}_v$ and $\mathbf{T}_v$ on the right-hand side of Eqs. (\ref{eq:lag_vel_modified}) and (\ref{eq:rot_vel_modified}), which is necessary for having the same order of convergence as the original method without virtual force as discussed in \citep{schwarz_Kempe_2015}.
Also, we follow the same initialization approach used in \citep{schwarz_Kempe_2015}.
The overall order of accuracy and convergence properties of the modified method and the temporal and spatial discretization errors are not changed by adding the virtual force, as mentioned in \citep{schwarz_Kempe_2015}, so a discussion on these topics is not repeated here.

\subsection{Lubrication force}
\label{sec:lubrication}
In numerical methods based on structured grids such as IBM, the flow field is not accurately resolved when the distance between the surface of particles becomes less than the grid spacing.
Therefore, the lubrication force is not completely resolved, which is important, especially in buoyant particles.
To resolve the lubrication force, it is necessary to use a fine grid, which results in very small time steps for the explicit scheme used here.
However, it has been argued in the literature that the details of the lubrication and collision model are only important when the trajectory of an individual particle is investigated, while the average statistics of large systems are not affected by these details \citep{akiki_2017,tavana_acta_2019}.

In this work, the lubrication force is modeled as \citep{simeonov_2012, sierakowski_2016}:
\begin{equation}
  \mathbf{F}_{lub}^{ij}=\begin{cases}
               0\;\;\;\;\;\;\;\;\;\;\;\;\;\;\;\;\;\;\;\;\;\;\;\;\;\;\;\;\;\;\;\;\;\;\;\;\;\;\;\;\;\;\;\;\;\;\;h < \epsilon_{col}\\
               \displaystyle{-6\pi\mu_f \frac{d_p^2}{16} \left( \frac{1}{h} - \frac{1}{\epsilon_{lub}} \right)\mathbf{U}_{ij}\cdot \mathbf{n}_{ij}}\;\;\;\;\epsilon_{col} < h < \epsilon_{lub}\\
               0\;\;\;\;\;\;\;\;\;\;\;\;\;\;\;\;\;\;\;\;\;\;\;\;\;\;\;\;\;\;\;\;\;\;\;\;\;\;\;\;\;\;\;\;\;\;\;h > \epsilon_{lub}
            \end{cases},
\end{equation}
where $\mathbf{F}_{lub}^{ij}$ is the lubrication force, $\mathbf{U}_{ij}$ is relative velocity between particle $i$ and $j$, $\mathbf{n}_{ij}$ is unit vector pointing out from the center of particle $i$ to the center of particle $j$, $h = \left| \mathbf{x}_i - \mathbf{x}_j \right| - d_p$ is the surface-to-surface distance between particles $i$ and $j$, $\epsilon_{lub}$ is the cutoff distance beyond which the lubrication force is negligible and $\epsilon_{col}$ has a nonzero positive value to prevent the singularity in the lubrication force as $h \rightarrow 0$.
Even with the inclusion of the lubrication force, some particles may collide with each other.
In this case, we use the same collision model already introduced, but with a small change that the collision starts when $h < \epsilon_{col}$, which prevents the lubrication force from becoming singular.
The parameters of lubrication force used in this study are $\epsilon_{lub}/d_p=0.5$ and $\epsilon_{col}/d_p=0.0003$, as suggested in the literature \citep{simeonov_2012,sierakowski_2016,akiki_2017}.

\section{Results and discussion}
\label{sec:results}
In this section, the simulation of a single buoyant particle is presented first, with the goal of the validation.
Then, the rise of two in-line buoyant particles is presented and compared with experimental results.
Finally, simulations of buoyant particle suspensions at different volume fractions are presented, and a drag law for buoyant particle suspensions is proposed based on the results of this part.

\subsection{Rise of a single buoyant particle}
\label{sec:single}
The goal of this subsection is to show that the implementation of the virtual force in PUReIBM is done correctly and to validate the numerical simulation.
Two different comparisons are made with other numerical and experimental works in the literature.
In the first case, the temporal evolution of particle velocity is studied, and results are compared with other numerical works.
Then, the drag force on a single buoyant particle is compared with the drag on a spherical bubble in contaminated liquid from an experiment.

\subsubsection{Temporal evolution}
\label{sec:single_time}
In this subsection, the motion of a single sphere ascending in a quiescent, viscous fluid under the action of gravity is simulated.
The input dimensionless parameters are Archimedes number $Ar$ and density ratio $\rho_p/\rho_f$.
The Archimedes number is defined as:
\begin{equation}
\label{eq:archimedes}
Ar = \frac{\rho_f |\rho_p - \rho_f|gd_p^3}{\mu_f^2}.
\end{equation}

The simulation is performed in a cuboidal domain with periodic boundary conditions in all directions.
The length of the domain in the direction of gravity is $L_x = 12.8d_p$, which is twice the length of the domain in other directions.
The particle and fluid are initially at rest and evolve under the action of gravity.
A mean pressure gradient is imposed on the system to oppose gravity and keep the mean fluid velocity zero, and then the particle starts to move due to the buoyant force.
Although a condition for $C_v^{min}$ was derived in Eq. (\ref{eq:cv_condition}) for the case of point particle simulations, other parameters affect $C_v^{min}$ in fully resolved simulations such as the grid resolution. 
For example, the lowest density ratio that can be simulated in PUReIBM (without using virtual force stabilization) is $\rho_p / \rho_f = 0.09$ with grid resolution per particle diameter $D_m = 20$ and $\rho_p / \rho_f = 0.07$ with $D_m = 30$.
The $C_v^{min}$ for these two cases are $0.15$ and $0.11$, respectively.
In our simulations, $C_v=0.15$ is used for the case of a single particle.

\begin{figure} [H]
\begin{centering}
\includegraphics[scale=0.5]{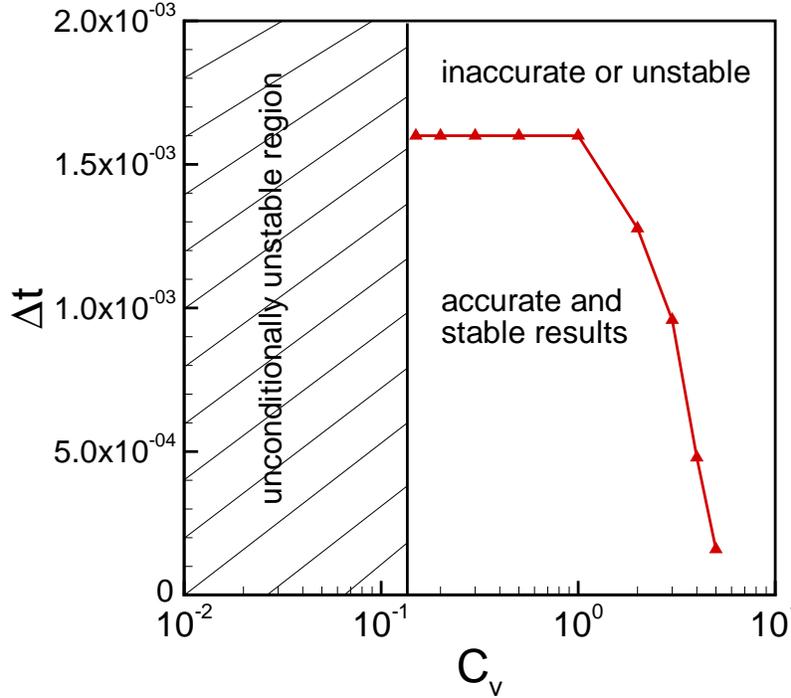}
\caption{Stability region for the simulation of a single rising particle for $D_m=20$. Simulations are unconditionally unstable for $C_v$ less than $C_v^{min}=0.15$ independent of the grid resolution and time step. The simulations are stable for $C_v > C_v^{min}$, however, to get accurate results it is necessary to limit the time step.}
\label{fig:stability}
\end{centering}
\end{figure}

Figure \ref{fig:stability} shows the region of stability for the simulation of a single rising particle with $Ar=1000$ and $D_m=20$.
Simulations are unconditionally unstable for $C_v$ less than $C_v^{min}$ independent of the grid resolution and time step.
This figure also shows that the simulations are stable for $C_v > C_v^{min}$, however, to get accurate results, it is necessary to limit the time step.
In the single rising particle simulation under gravity, the velocity at the beginning is very small, and using a constant $CFL = u_{max} {\Delta t}/{\Delta x}$ will result in a large time step.
Therefore a constant time step is used in PUReIBM with a condition on the maximum CFL number.
If the CFL number becomes larger than the maximum CFL number, then the time step is decreased.
For the case shown in Fig. \ref{fig:stability}, the CFL number does not become larger than the maximum CFL number, and hence the time step remains constant during the simulation.

To validate the result for the rise velocity of a single particle, it is compared with the numerical simulation using IBM developed by \citet{schwarz_Kempe_2015} and an implicit, highly-resolved spectral body-fitted method developed in Du{\v s}ek's group \citep{jenny_jcp_2004,jenny_jfm_2004}.
The results from Du{\v s}ek's group were received by \citet{schwarz_Kempe_2015} in private communication and are published in their work.

\begin{figure} [H]
\begin{centering}
\includegraphics[scale=0.5]{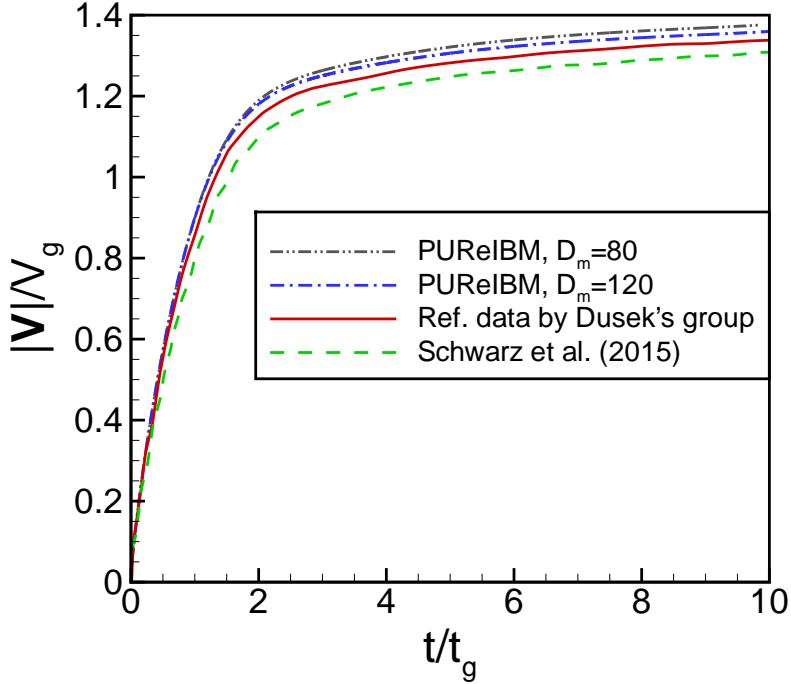}
\caption{Temporal evolution of a single rising particle for $Ar = 28900$ and $\rho_p/\rho_f = 0.001$.
Results of PUReIBM with two grid resolutions are compared to the reference body-fitted simulation of Du{\v s}ek's group \protect\citep{jenny_jcp_2004,jenny_jfm_2004} and the IBM simulation of \protect\citet{schwarz_Kempe_2015}.}
\label{fig:single_rise}
\end{centering}
\end{figure}

This comparison is shown in Fig. \ref{fig:single_rise} for $\rho_p/\rho_f = 0.001$ and $Ar = 28900$.
For the comparison, the gravitational velocity and time scale are utilized as reference values:
\begin{equation}
V_g = \sqrt{|\rho_p/\rho_f -1 |gd_p}, \;\;\;\;\;\;\;\;\; t_g = \sqrt{\frac{d_p}{|\rho_p/\rho_f -1 |g}}. \nonumber
\end{equation}

The results of PUReIBM are shown for two different grid resolutions in this figure.
At the early time, the PUReIBM results match very well with the reference result of Du{\v s}ek's group.
However, the terminal velocity is slightly different.
The difference decreases with increasing grid resolution, but convergence to the reference result is slow.
Similarly, the IBM results of \citet{schwarz_Kempe_2015} deviates from the reference terminal velocity.
In IBM, the no-slip and no-penetration velocity boundary conditions on the particle surface are imposed on Lagrangian marker points through the immersed boundary force
and then spread to the Cartesian grid using a regularized delta function.
In the implementation of IBM that is used by \citet{schwarz_Kempe_2015}, the Lagrangian marker points are on the surface of the particle, and the immersed boundary force is spread into the fluid domain, while in PUReIBM the Lagrangian marker points are inside the particle and the immersed boundary force is restricted to Eulerian grid points lying inside the sphere, while the fluid domain is uncontaminated by the immersed boundary force.
This could explain why the terminal velocity results of \citet{schwarz_Kempe_2015} are smaller than the reference data, and the results of PUReIBM are larger.

Note that only the beginning of the acceleration phase is considered for comparison in Fig. \ref{fig:single_rise}, which is before the particle path shows instability \citep{jenny_jfm_2004}.
When the Reynolds number is higher than a critical value, and the particle density is much smaller than that of the surrounding fluid, the particle motion is spiral, and the drag coefficient is almost constant.
At Reynolds number below the critical value, the drag coefficient follows the standard drag curve and the trajectory is linear\citep{karamanev_1996,veldhuis_2009,jenny_jcp_2004,jenny_jfm_2004, horowitz_williamson_2010,rahmani_2014}.
This behavior is explained as the mechanical inertia of the particle becomes small enough for the wake to induce rotation of the particle, thus creating a spiral trajectory \citep{karamanev_1996}.
In addition, for light particles, the dominant inertial force is the added mass from the attached fluid, which accelerates with the particle.
This effect is shown recently to play an important role in the dynamics of buoyant particles suspensions \citep{tavana_acta_2019}.

Another interesting test case for buoyant particles is the rise of a buoyant particle in an inclined channel.
In this problem, the particle rises in an inclined channel due to buoyancy and travels alongside the right wall of the domain.
\citet{lomholt_2002} have performed an experiment for this case for a particle with $\rho_p/\rho_f=0.97$.
However, PUReIBM works properly, even without the virtual force stabilization technique for this density ratio.
In other words, the results with and without virtual force stabilization would be the same for $\rho_p/\rho_f=0.97$.
Therefore we have presented this test case in \ref{sec:appB}.

\subsubsection{Drag coefficient of a single buoyant particle}
\label{sec:single_drag}
As another validation, the results of PUReIBM for the drag coefficient of a single rising particle are compared with the experimental results of \citep{dijkhuizen_2010b}.
The experimental work reports the drag of nearly spherical gaseous bubbles (aspect ratio $E > 0.95$) in tap (contaminated) water.
As explained already, buoyant particles are a good approximation for bubbles in contaminated liquid, so this comparison is valid.

Additionally, the results are compared with the drag correlation for a single bubble in an unbounded medium of contaminated liquid proposed by \citet{tomiyama_1998}.
This correlation is expressed as:
\begin{equation}
\label{eq:cd}
C_{d,0} = \text{max} \left[ \frac{24}{Re} \left(1+0.15 Re^{0.687}\right),\frac{8}{3} \frac{Eo}{Eo+4} \right],
\end{equation}
where $Eo$ is E{\"o}tv{\"o}s number which represents the ratio between buoyancy and surface tension forces and is defined as,
\begin{equation}
\label{eq:Eo}
Eo = \frac{g \left( \rho_f - \rho_p \right) d_p^2}{\sigma},
\end{equation}
with $\sigma$ being the surface tension.
Note that $Eo$ for rigid particles corresponds to $\sigma \rightarrow \infty$, so $Eo$ is zero in our simulations which means Eq. (\ref{eq:cd}) reduces to the famous Schiller--Naumann drag coefficient. 
In Eq. \ref{eq:cd}, Reynolds number is defined with the rise velocity of the particle as:
\begin{equation}
  \label{eq:Re_def}
  Re = \frac{\rho_f \abs{\mathbf{V}} d_p}{\mu_f}.
\end{equation}

\begin{figure} [H]
\begin{centering}
\includegraphics[scale=0.5]{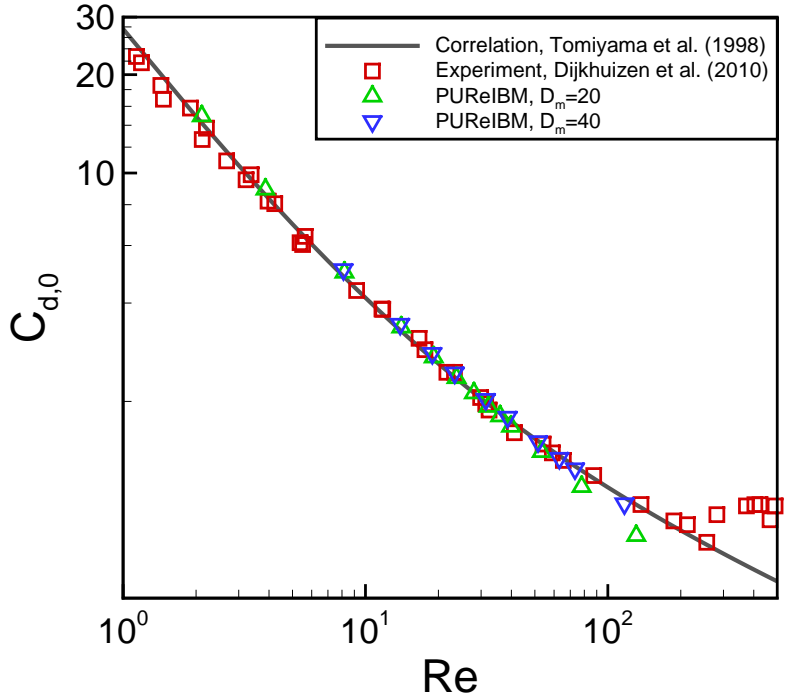}
\caption{Drag coefficient versus Reynolds number from simulation of a single buoyant particle in PUReIBM compared with experimental results of a nearly spherical bubble in contaminated liquid \protect\citep{dijkhuizen_2010b} and drag correlation for a single bubble in contaminated liquid \protect\citep{tomiyama_1998}.}
\label{fig:single_exp}
\end{centering}
\end{figure}

The simulation setup is similar to the previous case (section \ref{sec:single_time}) with $\rho_p/\rho_f = 0.001$ and different $Ar$ to achieve different Reynolds number defined by the terminal velocity.
Figure \ref{fig:single_exp} shows the comparison of PUReIBM results with experimental results and the drag correlation in Eq. (\ref{eq:cd}).
The results of PUReIBM are presented for two grid resolutions, and it is clear that the finer grid gives the correct results at a higher Reynolds number.
In general, the results of PUReIBM match very well with experiments and the correlation.

\subsection{Rise of two buoyant particles}
\label{sec:two}
The next simulation that is presented is for the rise of two in-line particles.
The particles rising in-line is a specific but typical case where mutual interactions between particles are evident.
In this particular case, the rise velocity of the trailing particle is affected (and increased) by the wake of the leading particle. 
Finally, the trailing particle reaches the leading particle.

\begin{figure} [H]
\begin{centering}
  \subfigure[]{ \includegraphics[clip, width=57mm]{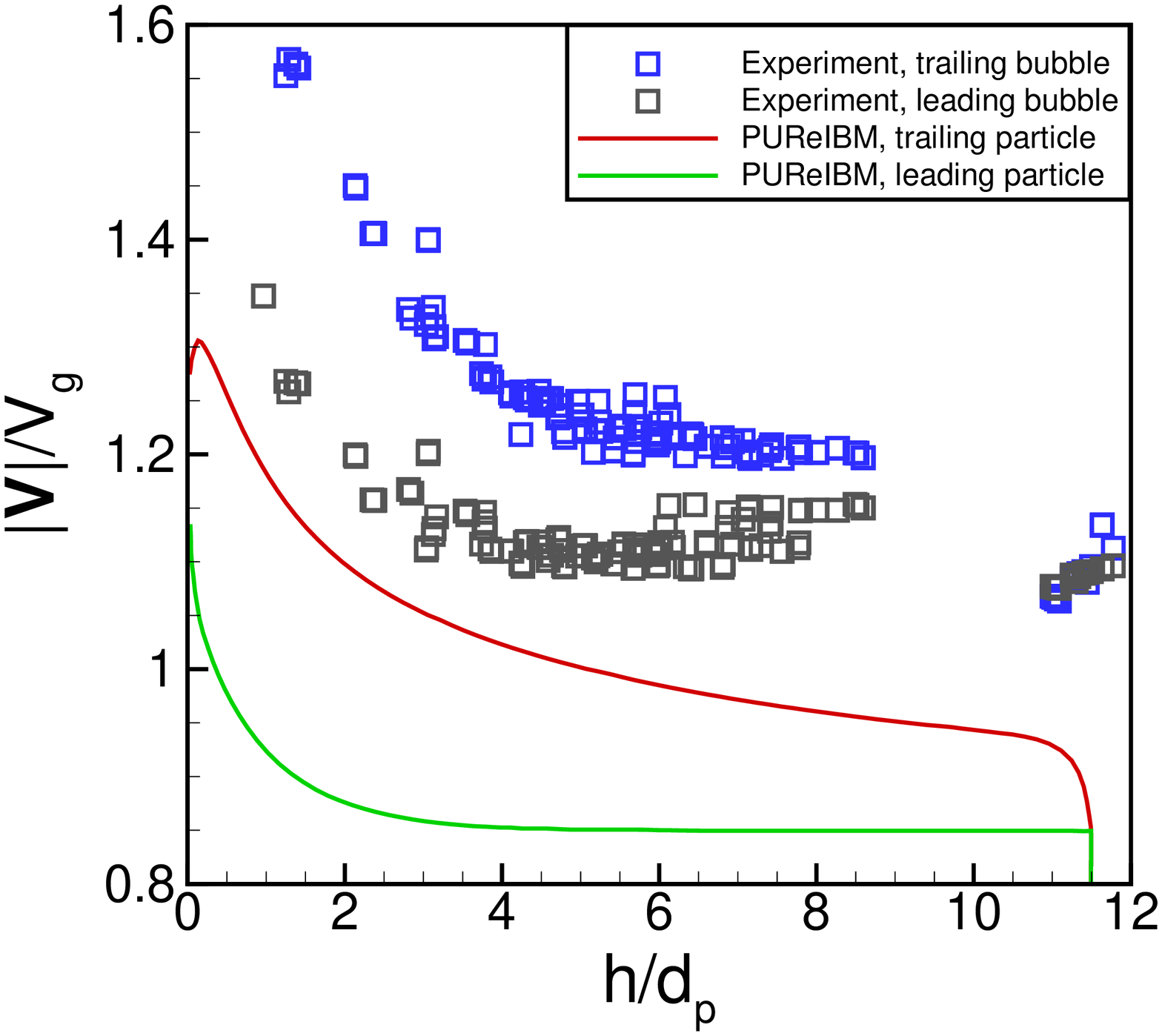} \label{fig:rise1}}
  \subfigure[]{ \includegraphics[clip, width=57mm]{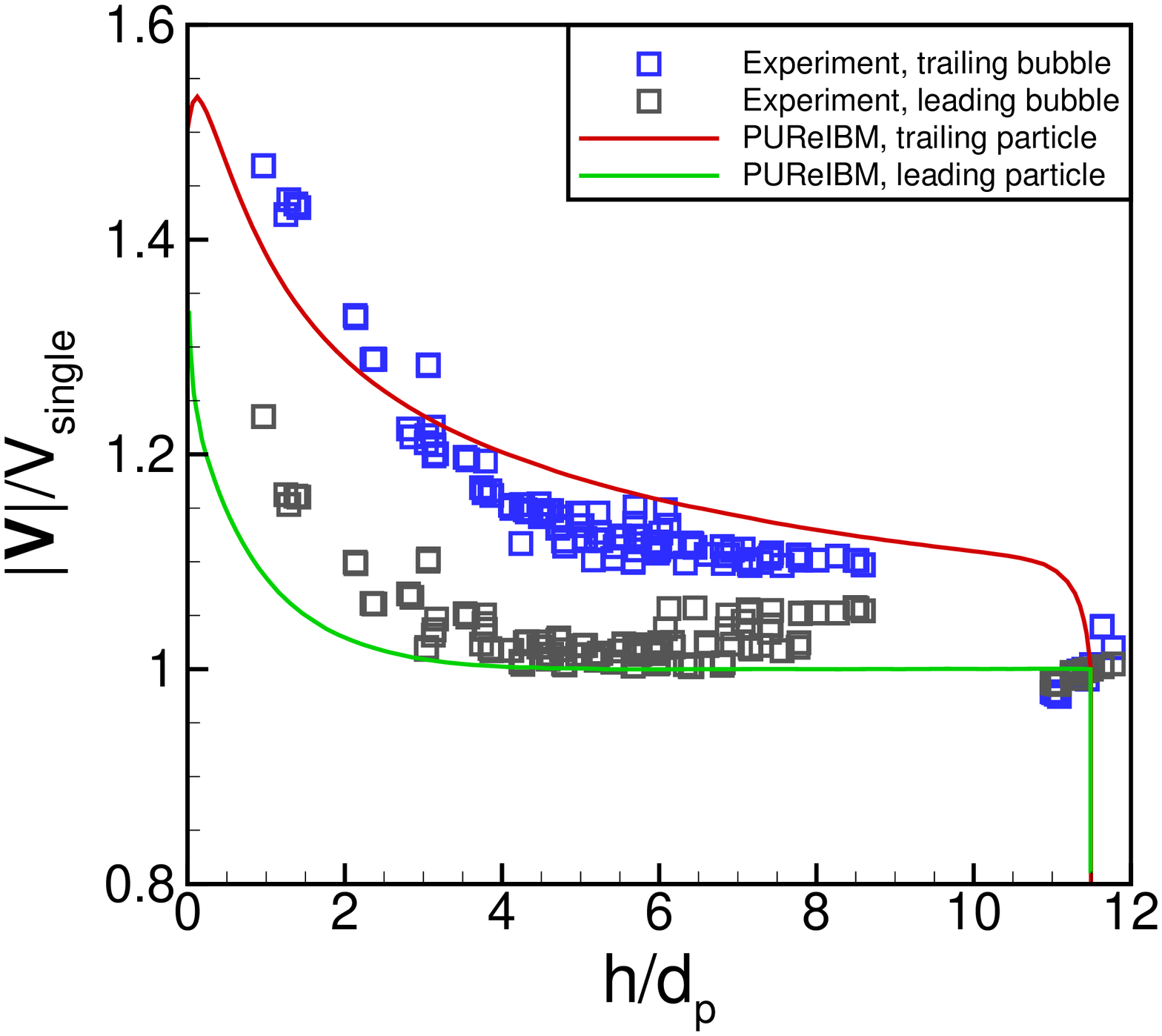} \label{fig:rise2}}
  \caption{Rise velocity of two in-line particles from simulation in PUReIBM compared with experimental results of two in-line bubbles in distilled water \protect\citep{katz_1996}.
  \subref{fig:rise1} Rise velocity is scaled by the gravitational velocity.
  \subref{fig:rise2} Rise velocity is scaled by the rise velocity of a single particle/bubble.}
\label{fig:two2}
\end{centering}
\end{figure}

The simulation is again performed in a cuboidal domain with periodic boundary conditions in all directions.
The length of the domain in the direction of gravity is $L_x = 89.6d_p$, which is $14$ times the length of the domain in other directions.
Similar to the single-particle case, the particles rise due to the buoyant force.
The simulation is performed for a case with $Ar = 1700$, $\rho_p/\rho_f = 0.001$, $C_v=0.15$, and the initial surface-to-surface distance between the particles is $h_0/d_p=11.6$.
Figure \ref{fig:rise1} shows the rise velocity of particles versus surface-to-surface distance between them.
The experimental results of \citet{katz_1996}, shown in this figure, are for a similar case but bubbles in distilled water.
For both the numerical and experimental cases, the Reynolds number based on the terminal velocity of a single particle/bubble is $35.4$.
Although the trend of the rise velocity in both cases is similar, they do not match.
One reason for this is that the results of PUReIBM represent bubbles in contaminated liquid, while the experimental results are for bubbles in clean liquid.
Note that \citet{katz_1996} used commercially available distilled water and not highly purified liquid, which could also be considered a partially contaminated liquid.
Nevertheless, it is known that the drag force, and consequently, the rise velocity of bubbles in clean, partially contaminated, or contaminated is different.
To have a better comparison, the rise velocity scaled by the rise velocity of a single particle/bubble is plotted in Fig. \ref{fig:rise2}.
With this scaling, it is easier to compare the relative velocity between the leading and trailing particle/bubble as a function of the distance between them since we have now a common basis for comparison in both cases ($\abs{V}/V_{single}=1$ at $h/d_p=11.6$).
This figure shows that the behavior of bubbles in contaminated or clean liquid is comparable if a proper scaling is used.

\subsection{Simulation of buoyant particle suspensions}
\label{sec:swarm}
The main goal of this paper is to perform simulations of buoyant particles at high volume fractions.
In this section, the results of simulations for buoyant particle suspensions at volume fraction $0.1$ to $0.4$ are presented, and in particular, the drag force on the suspension is discussed.

\subsubsection{Problem setup}
\label{sec:swarm_setup}
In this part, the simulations are performed in a cubic domain with periodic boundary conditions.
The initial positions of the particles are obtained following elastic collisions (in the absence of interstitial fluid), starting from a lattice arrangement with a Maxwellian velocity distribution.
The particles and fluid are initially at rest and evolve under the action of gravity.
A mean pressure gradient is imposed on the system to oppose the effect of gravity and keep the mean fluid velocity zero, and then the particle starts to move due to the buoyant force.
The simulations are carried out until the mean particle velocity reaches a statistically stationary state.
In our simulations, the mean drag on particles or mean particle velocity is computed by averaging over all particles and then ensemble-averaging over different particle configurations.
For each case, five independent realizations (corresponding to a specified initial particle configuration) are simulated in this study.

We have performed simulations for five different $Ar$ and four different $\phi$.
The salient numerical and physical parameters used in the simulations are reported in Table \ref{tab:num_params}.
Note that the Reynolds number based on the rise velocity of the suspension $Re$ is the outputs of the simulations.
Note that an increase in volume fraction decreases the rise velocity (and the Reynolds number), as indicated in Table \ref{tab:num_params}.
In this Table, $Re_0$ is the Reynolds number of a single particle at each $Ar$ number which is calculated by balancing the drag and buoyancy forces on a single particle which results in $Re^2_0 = 4\;Ar/\fn{3\;C_{d,0}}$, where we have used Eq. (\ref{eq:cd}) for the drag coefficient.
For all cases density ratio is $\rho_p/\rho_f=0.001$ and number of particle is $N_p=200$.
We have decreased the length of the domain by reducing the volume fraction to keep the number of particles fixed.
The length of the domain is $L/d_p = 10.08, 8.06, 7.05, 6.4$ for $\phi = 0.1, 0.2, 0.3, 0.4$, respectively.
The length of the domain is chosen to ensure that the two-point correlation functions in the fluid phase decay to zero within the box length \citep{tenneti_ijmf_2011}.
The grid resolution used in this study is sufficient to obtain converged results for the mean drag and second moments of particle and fluid velocities.

\begin{table} [H]
\centering
\caption{The numerical and physical parameters of the simulations: Archimedes number $Ar$, Reynolds number of a single particle $Re_{0}$, volume fraction of particles $\phi$, the number of grid cells across the diameter of a particle $d_p/\Delta x$, Reynolds number of the suspensions $Re$. For all cases $N_p=200$ and $\rho_p/\rho_f=0.001$.}
\label{tab:num_params}     
\begin{tabular}{c c c c c} 
\hline\noalign{\smallskip}
$Ar$ & $Re_{0}$ & $\phi$ & $d_p/\Delta x$ & $Re$ \\
\noalign{\smallskip}
\hline\noalign{\smallskip}
$5 \times 10^3$ & $72.33$ & $0.1, 0.2, 0.3, 0.4$ & $30, 30, 30, 30$ & $48.66, 37.96, 29.36, 21.13$ \\
\hline\noalign{\smallskip}
$1 \times 10^4$ & $113.85$ & $0.1, 0.2, 0.3, 0.4$ & $30, 30, 30, 30$ & $78.28, 61.13, 47.16, 35.52$ \\
\hline\noalign{\smallskip}
$2 \times 10^4$ & $177.43$ & $0.1, 0.2, 0.3, 0.4$ & $30, 30, 30, 30$ & $120.28, 95.81, 76.19, 57.14$ \\
\hline\noalign{\smallskip}
$5 \times 10^4$ & $315.27$ & $0.1, 0.2, 0.3, 0.4$ & $40, 40, 40, 40$ & $206.21, 161.21, 135.17, 107.87$ \\
\hline\noalign{\smallskip}
$1 \times 10^5$ & $483.82$ & $0.1, 0.2, 0.3, 0.4$ & $50, 50, 40, 40$ & $309.37, 256.88, 209.22, 161.25$ \\
\noalign{\smallskip}
\hline
\end{tabular}
\end{table}

According to Eq. (\ref{eq:cv_condition}), $C_v^{min}$ is related to the added mass coefficient.
Many studies have shown that added mass coefficient increases with volume fraction \citep{zuber_1964,wijngaarden_jeffrey_1976,biesheuvel_spoelstra_1989, sankar_2002,beguin_2016}.
\citet{spelt_sangani_1997} also reported the same dependence, however, they included the effect of microstructure on the added mass through the velocity fluctuations of bubbles.
On the other hand, \citet{simcik_2008} and \citet{simcik_2013} have reported that the added mass coefficient can decrease or increase with volume fraction depending on the shape of the computational domain.
In our simulations, we have to increase  $C_v^{min}$ with volume fraction to get stable results, which indicates that the added mass coefficient increases with volume fraction.

In PUReIBM, $C_v^{min}$ is found from numerical experiments to increase from $0.18$ for $\phi=0.1$ to $0.22$ for $\phi=0.4$.
Using the correlation by \citet{zuber_1964}, $C_{am}$ increases from $0.66$ for $\phi=0.1$ to $C_{am}=1.5$ for $\phi=0.4$.
It shows that the growth of $C_v^{min}$ with volume fraction is slower than the growth of $C_{am}$.
In this study, $C_v = 0.25$ is used for all volume fractions in the simulation of particle suspensions.
It is also important to mention that there is no added mass effect in an average sense since the simulations reach a statistically stationary state; however, each individual particle experiences the added mass effect.
Therefore, using the virtual force stabilization technique is necessary at any stage of the simulations.

\subsubsection{Drag coefficient of the buoyant particle suspensions}
\label{sec:swarm_drag}
The results presented in this section are the drag coefficient $C_d$ of the buoyant particle suspensions obtained from PUReIBM, which are compared with experimental and numerical results for bubbly flows.
The drag coefficient for the suspension is scaled with the drag coefficient of a single particle $C_{d,0}$ at the same Archimedes number multiplied with $\epf$.
As a result the scaled drag coefficient is equal to the squared of the ratio of the rise velocity of a single particle to the rise velocity of the suspensions, i.e. \citep{roghair_2011}:
\begin{equation}
\label{eq:cd_swarm}
\frac{C_d}{C_{d,0} \epf} = \left( \frac{\abs{\mathbf{V}_0}}{\abs{\langle \mathbf{V} \rangle }} \right)^2 = \left( \frac{Re_{0}}{Re} \right)^2 = f.
\end{equation}
In general, $f$ could be a function of $Re$, $\phi$, and $Eo$.
However, if $C_{d,0}$ already incorporates the dependencies on $Re$ and $Eo$, the function $f$ would only depend on $\phi$.
Similarly, we can say that the comparison of $f$ for clean and contaminated bubbles is only valid if the effect of contamination is considered in $C_{d,0}$. 
In other words, the drag of bubbles in clean and contaminated liquid is not comparable unless a proper scaling is used, as discussed in Section \ref{sec:two}.
For this reason, different studies have used $C_{d,0}$ from different correlations or measurements in Eq. (\ref{eq:cd_swarm}) for reporting $f$.
In this work, Eq. (\ref{eq:cd}) is used for $C_{d,0}$ for scaling.

All the experimental correlations used in this paper for comparison only depend on the volume fraction (see Table \ref{tab:cd_swarm_exp}).
This means that experimental studies verify that the effects of $Re$ and $Eo$ are fully accounted for in $C_{d,0}$.
The form of the correlation by \citet{bridge_1964} is inspired by \citet{richardson_zaki_1954}.
\citet{rusche_issa_2000} used data from different experimental works in the literature to propose their correlation.
Their correlation is a rough fit to a lot of experimental data with large deviations for $ 5 \times 10^4 < Ar < 1 \times 10^7$.
The highest Archimedes number in our simulations is $ Ar = 1 \times 10^5 $.
\citet{garnier_2002} performed their experiments in a highly controlled environment with a uniform swarm of monodisperse bubbles without recirculating motions in the liquid phase, and their correlation is also verified in experiments by \citet{guet_2004}.
Their correlation comes from experiments with $ 300 < Re < 500$.

\begin{table} [H]
\centering
\caption{Experimental correlations for the scaled drag coefficient, i.e., function $f$ in Eq. (\ref{eq:cd_swarm}).}
\label{tab:cd_swarm_exp}     
\begin{tabular}{c c c} 
\hline\noalign{\smallskip}
Correlation & Condition & Reference \\
\noalign{\smallskip}
\hline\noalign{\smallskip}
\rule{0pt}{20pt}$\left[ \epf^{1.39} \right]^{-2}$ & $ \phi < 0.2$ & \citet{bridge_1964} \\
\rule{0pt}{20pt}$\left[ \exp{\left(3.64 \phi \right)} + \phi^{0.864} \right] \epf^{-1}$ & $ \phi < 0.45$ & \citet{rusche_issa_2000} \\
\rule{0pt}{20pt}$\left[ 1 - \phi^{1/3} \right]^{-2}$ & $ \phi < 0.3 $ & \citet{garnier_2002} \\
\noalign{\smallskip}
\hline
\end{tabular}
\end{table}

The numerical correlations by \citet{roghair_2011,roghair_2013} are obtained from FR-DNS using the front-tracking method (FTM) and depend on both volume fraction and E{\"o}tv{\"o}s number (see Table \ref{tab:cd_swarm_num}).
Based on their work, it is concluded that only the effect of $Re$ is incorporated in $C_{d,0}$ and not $Eo$.
Our simulations of buoyant particles correspond to spherical bubbles in contaminated liquid at $Eo=0$.
So, our results in this section are compared with Roghair's second and third correlations since they are developed for smaller values of $Eo$.

\begin{table} [H]
\centering
\caption{Numerical correlations by \protect\citet{roghair_2011,roghair_2013} for the scaled drag coefficient, i.e., function $f$ in Eq. (\ref{eq:cd_swarm}).}
\label{tab:cd_swarm_num}     
\begin{tabular}{l c c} 
\hline\noalign{\smallskip}
Correlation & Condition & Reference \\
\noalign{\smallskip}
\hline\noalign{\smallskip}
\rule{0pt}{25pt}$\displaystyle{1 + \left( \frac{18}{Eo} \right) \phi}$ & \begin{tabular}{@{} c @{}} wobbling bubbles \\ $ 1.2 < Eo < 4.8 $ \\ $ 0.05 < \phi < 0.45$ \end{tabular} & \citet{roghair_2011} \\
\rule{0pt}{25pt}$\displaystyle{1 + \left( \frac{22}{Eo+0.4} \right) \phi}$ &  \begin{tabular}{@{} c @{}} spherical/wobbling bubbles \\ $ 0.134 < Eo < 4.8 $ \\ $ 0.05 < \phi < 0.4$ \end{tabular} & \citet{roghair_2013} \\ 
\rule{0pt}{25pt}$\displaystyle{1 + \left( \frac{6.612 Eo + 2.023}{Eo} \right) \phi} $ & \begin{tabular}{@{} c @{}} spherical/ellipsoidal bubbles \\ $ 0.5 < Eo < 2 $ \\ $ 0.05 < \phi < 0.15$ \end{tabular} & \citet{roghair_2013} \\
\noalign{\smallskip}
\hline
\end{tabular}
\end{table}

\begin{figure} [H]
\begin{centering}
\includegraphics[scale=0.7]{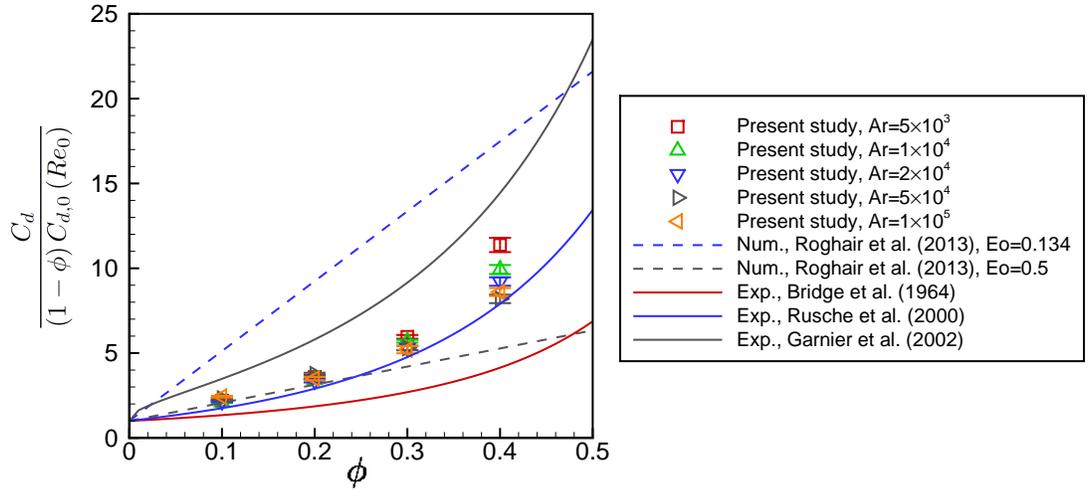}
\caption{Comparison of drag coefficient of the suspension scaled with $C_{d,0} \fn{Re_0}$ obtained from PUReIBM with different drag correlations.
Solid lines are the correlations obtained from different experiments (see Table \ref{tab:cd_swarm_exp}).
Dashed lines are the correlations obtained from simulations using FTM \protect\citep{roghair_2013} for two different values of $Eo$ (see Table \ref{tab:cd_swarm_num}).
Symbols are the scaled drag coefficients obtained from PUReIBM (present study) for different values of $Ar$.
The error bars represent $95 \%$ confidence intervals obtained from five independent realizations for each case.}
\label{fig:drag_swarm}
\end{centering}
\end{figure}

Figure \ref{fig:drag_swarm} compares the results for the scaled drag coefficient in buoyant particle suspensions and bubbly flows.
The correlations are extended to higher volume fractions if the range for which they are proposed covers a smaller range.
Our results in Fig. \ref{fig:drag_swarm} show that the scaled drag coefficient depends on both Archimedes number (or equivalently Reynolds number) and volume fraction.
However, there is a trend in the results, which shows that with increasing $Ar$ the scaled drag becomes only a function of volume fraction.
In other words, the data for higher $Ar$ collapse to a single line.
Nevertheless, it is interesting to note that the results of PUReIBM, similar to the experimental correlations, show a nonlinear dependence of drag coefficient on volume fraction.
In contrast, all correlations of clean bubbles using FTM have linear dependence (see Table \ref{tab:cd_swarm_num}). 
Although \citet{roghair_2011} do not provide the reason behind the linear nature of their correlations, they raise four possible issues in their work including the effect of 1) contamination, 2) coalescence and breakup, 3) normalization with $C_{d,0}$ which comes from different correlations or measurements for each case, and 4) having a smaller computational domain in comparison to the large domain of experiments.

Of these, two limitations are addressed in this work.
First, the bubbles in this study are contaminated bubbles, and the results of experimental works might also have some level of contamination.
Secondly, the FR-DNS results from FTM by \citet{roghair_2011,roghair_2013} can only take local gas fractions into account, since the computational domain is small compared to the physical domain typically used in experiments.
In fact, \citet{roghair_2011,roghair_2013} have between $16$ to $32$ bubbles for different simulations while $200$ particles is used for each case in PUReIBM.
Therefore, it is concluded that PUReIBM results have a similar trend to experiments since they are performed in larger domains with more particles.

It should also be mentioned that \citet{simonnet_2007} developed a drag correlation using local volume fraction definition in their experiments, but their correlation predicts that the drag coefficient increases very slowly up to volume fraction $15 \%$ and then decreases with increasing volume fraction.
The main reason for the different behavior they obtained is that the bubbles in their experiment are large $ \left( d_b > 7 mm \right) $, and because of this, their correlation is not presented here.

\subsubsection{Drag correlation of the suspension}
\label{sec:drag_correlation}
In general, fully resolved simulations are a useful tool for model development for macro-scale simulations \citep{tenneti_arfm_2014}.
For instance, drag force obtained from fully resolved simulation has been already used to develop a model for interphase momentum transfer in two-fluid equations \citep{hill_koch_2001a,hill_koch_2001b,vanderHoef_2005,beetstra_2007,tenneti_ijmf_2011}.
Similarly, we can propose a correlation for the drag coefficient using our data presented in Fig. \ref{fig:drag_swarm}.
However, a correlation based on this data set will be a function of both $Ar$ (or $Re$) and $\phi$.
Interestingly, if we scale the drag coefficient of the suspension with the drag coefficient of a single particle at the same Reynolds number of the suspension (instead of the drag coefficient of a single particle at the same Archimedes number of the suspension which results in $Re_0$ reported in Table \ref{tab:num_params}),  all data collapse to a single line.
In other words, in this new scaling, we calculate $C_{d,0}$ from Eq. (\ref{eq:cd}) with ${Re_m}$ instead of ${Re_0}$ where $Re_m$ is the Reynolds number based on the superficial velocity of the suspensions:
\begin{equation}
\label{eq:Rem}
Re_m = \fn{1-\phi} Re = \frac{\rho_f \fn{1-\phi} \abs{ \langle \mathbf{V} \rangle} d_p}{\mu_f}.
\end{equation}

\begin{figure} [H]
\begin{centering}
\includegraphics[scale=0.5]{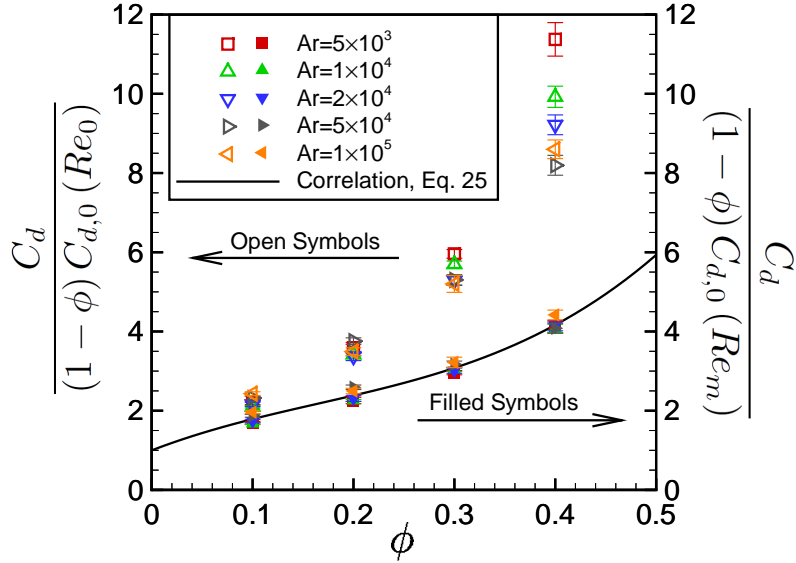}
\caption{Drag coefficient of the suspension scaled with $C_{d,0} \fn{Re_0}$ (empty symbols) and $C_{d,0} \fn{Re_m}$ (filled symbols) obtained from PUReIBM (present study) for different values of $Ar$.
Solid line is the correlation developed (Eq. \ref{eq:cd_correlation}) based on scaling with $C_{d,0} \fn{Re_m}$.
The error bars represent $95 \%$ confidence intervals obtained from five independent realizations for each case.}
\label{fig:drag_correlation}
\end{centering}
\end{figure}

The results with both scaling are shown in Fig. \ref{fig:drag_correlation}.
Using the Reynolds number of the suspension in correlations of drag law is common and several other correlations for gas--solid flows (fixed bed or freely evolving) are proposed based on the Reynolds number with mean slip velocity in the suspension\citep{wen_yu_1966,beetstra_2007,tenneti_ijmf_2011,zaidi_2014,bonger_2015,tang_2015, tang_free_2016,zaidi_2018}.
Since our results with this new scaling are not affected by $Ar$ significantly, the scaled drag coefficient is modeled here as only a function of volume fraction.
We propose the following correlation by curve fitting (with $R^2 = 0.9903$):
\begin{equation}
\label{eq:cd_correlation}
\frac{C_d}{\epf C_{d,0} \fn{Re_m}} = 48.51 \phi^3 - 24.15 \phi^2 + 9.81 \phi + 1.
\end{equation}
Note that in developing this correlation, we have used the fact that this ratio should be unity at $\phi=0$.

FR-DNS can also be used to develop stochastic or deterministic models of acceleration in point-particle simulations \citep{tenneti_jfm_2016,akiki_jackson_balachandar_2017,akiki_2017,esteghamatian_2018} or for model development in Euler--Lagrange simulation where Eulerian equations are volume filtered \citep{capecelatro_2013}.
This work opens the door to the development of such models for buoyant particles in future studies.

\section{Conclusions}
\label{sec:conclusion}
In this work, a FR-DNS solver based on PUReIBM is developed for simulation of buoyant particles with a density ratio as small as $0.001$ and for a range of volume fraction up to $40 \%$.
It is explained that when the mass of a particle is smaller than the added mass induced by the surrounding fluid, explicit numerical methods are unstable.
To stabilize the method, the virtual force technique introduced by \citet{schwarz_Kempe_2015} is extended.
It is shown that the virtual force constant $C_v$ has a lower limit for having stable simulations, which depends on the density ratio and the added mass coefficient.
Since the added mass coefficient increases with an increase in the volume fraction of particles, it is concluded that $C_v$ should also increase in the case of buoyant particle suspensions when compared to the single-particle case.

Simulations of rigid buoyant particles are performed, which are considered a good approximation for bubbles in contaminated liquid.
The results from simulations of a single buoyant particle at different Archimedes number are presented and compared with the numerical and experimental reference data to validate the implementation of the virtual force in PUReIBM.
Then two in-line rising buoyant particles are simulated and compared with experimental results.
Finally, simulations of freely evolving buoyant particle suspensions are presented.
The scaled drag coefficient of particle suspensions at different volume fractions are compared with numerical and experimental correlations of bubbly flows from the literature.
It is shown the increase in the scaled drag coefficient of buoyant particles with volume fraction from FR-DNS using PUReIBM matches the nonlinear trend in experimental data of bubbly flows, while the correlations from FR-DNS using FTM by \citet{roghair_2011,roghair_2013} predict a linear increase.
They hypothesized that this linear dependence could be to the fact that their bubbles are in clean liquid, and their simulations are performed for a small domain with $16$ up to $32$ bubbles in it.
It is shown here that considering a larger domain with $200$ buoyant particles (bubbles in contaminated liquid) predicts the nonlinear behavior seen in experimental studies.
Finally, it is shown that scaling the drag coefficient of the suspension with the drag coefficient of a single particle at the same Reynolds number of the suspension is only a function of the volume fraction and a drag correlation is presented based on data using this scaling.

\section*{Acknowledgments}
This material is based on work supported by the National Science Foundation under Grant No. 1438143.

\begin{appendix}
\section{Numerical calculation of virtual force and torque}
\label{sec:appA}
As explained in section \ref{sec:pureibm_virtualforce}, one needs to numerically calculate $\mathbf{F}_v$ and $\mathbf{T}_v$ on the right-hand side of Eqs. (\ref{eq:lag_vel_modified}) and (\ref{eq:rot_vel_modified}) before numerically integrate them to calculate $\mathbf{V}$ and $\mathbf{\Omega}$ in time.
In this Appendix, we explain how it is done for virtual force.
The same approach is used for virtual torque.
Eq. (\ref{eq:lag_vel_modified}) can be re-written as:
\begin{equation}
	\label{eq:simple_lag_vel_modified}
	\frac{d \mathbf{V}}{dt} = \alpha_f \boldsymbol{\xi}_f + \alpha_v \boldsymbol{\xi}_v,
\end{equation}
where $\displaystyle{\alpha_f =\frac{1}{\fn{\rho_p + C_v \rho_f} V_p}}$, $\displaystyle{\boldsymbol{\xi}_f = \mathbf{F}_B + \mathbf{F}_h + \sum_{\substack{{j=1}\\{j \neq i}}}^{N_p} \mathbf{F}_{c}^{(ij)}}$, $\displaystyle{\alpha_v =\frac{C_v \rho_f}{\rho_p + C_v \rho_f}}$, $\displaystyle{\boldsymbol{\xi}_v = \frac{d \mathbf{V}}{dt}}$.

The following method which is a combination of the Adams-Bashforth predictor-corrector scheme and the trapezoidal rule is used to approximate $\boldsymbol{\xi}_v$ and then update the velocity:
\begin{align}
	\label{eq:AB2_predict}
	\boldsymbol{\xi}_v^n = \frac{3\mathbf{V}^n-4\mathbf{V}^{n-1}+\mathbf{V}^{n-2}}{2\Delta t} \nonumber \\ 
	\tilde{\boldsymbol{\xi}}^n = \alpha_f \boldsymbol{\xi}_f^n + \alpha_v \boldsymbol{\xi}_v^n \nonumber \\
	\mathbf{V}^{n+1}_{pred} = \mathbf{V}^{n} + \frac{\Delta{t}}{2} \left( 3\tilde{\boldsymbol{\xi}}^n -\tilde{\boldsymbol{\xi}}^{n-1} \right) \nonumber \\
	\boldsymbol{\xi}_{v,pred}^{n+1} = \frac{3\mathbf{V}^{n+1}_{pred}-4\mathbf{V}^{n}+\mathbf{V}^{n-1}}{2\Delta t} \nonumber \\
	\boldsymbol{\xi}_v^{n+\frac{1}{2}} = \frac{1}{2} \left( \boldsymbol{\xi}^n_v + \boldsymbol{\xi}^{n+1}_{v,pred} \right) \nonumber \\
	\mathbf{V}^{n+1} = \mathbf{V}^{n} + \Delta{t} \left( \alpha_f \boldsymbol{\xi}_f^n + \alpha_a \boldsymbol{\xi}_v^{n+\frac{1}{2}} \right)
\end{align}
 
\section{Rise of a single buoyant particle in an inclined channel}
\label{sec:appB}
Here the rise of a buoyant particle in an inclined channel is simulated, which corresponds to the experiment by \citet{lomholt_2002}.
The particle rises in this inclined channel due to buoyancy and travels alongside the right wall of the domain.
The simulation is performed for a particle with $\rho_p/\rho_f=0.97$, which is the same as experimental work.
For this density ratio, PUReIBM works properly even without the virtual force stabilization technique.
As mentioned earlier, the lowest density ratio that PUReIBM becomes unstable without using virtual force is $\rho_p/\rho_f = 0.07$.
In other words, the results with and without virtual force would be the same for $\rho_p/\rho_f=0.97$.
Since this test case is a general validation for PUReIBM, we have presented it in this appendix.
The experiment is performed for $Re_p^{Stokes}=13.6$ which is the Reynolds number based on the Stokes settling velocity $W$ and is defined as
\begin{equation}
	\label{eq:Re_p_stokes}
	Re_p^{Stokes} = \frac{W d_p}{\nu} = \frac{d_p^3}{18 \nu} |\frac{\rho_p}{\rho_f} - 1|g.
\end{equation}
It is clear from Eq. (\ref{eq:archimedes}) that $Ar = 18 \; Re_p^{Stokes}$.
For comparison, we have performed the simulation at $Ar=244.8$.
The channel is inclined at an angle of $8.23^{\circ}$ with the vertical.
Numerically, this is simulated by adding components of gravitational forces in the horizontal and vertical directions. 
The computational domain consists of a rectangular box $L/d_p = \fn{20,5,40}$.
The grid is Cartesian and uniform over the domain with $d_p/\Delta x = 20$. 
The particle is injected at $\bld{x}_0/d_p = \fn{10,1.8,-0.5}$.

The numerical method used in PUReIBM to solve Navier--Stokes equations is a pseudo-spectral method with Fourier basis functions.
This means that we can only impose periodic boundary conditions in our code (and not the no-slip boundary condition for the wall).
However, we can simulate a wall (imposing no-slip boundary condition on a plane) using the immersed boundary approach itself.
We have done this already for single-phase flow \citep{mehrabadi_jfm_2015} and heat transfer \citet{tenneti_ijhmt_2013} in a duct.
Therefore, the periodic boundary condition is imposed in all directions, and a wall is generated in $xz-$plane at $y/d_p=5$ using the immersed boundary method.
Although we only generate the wall at one location, due to the periodic boundary condition, this will result in the creation of a channel.

\begin{figure} [H]
\begin{centering}
  \subfigure[]{ \includegraphics[clip, width=37mm]{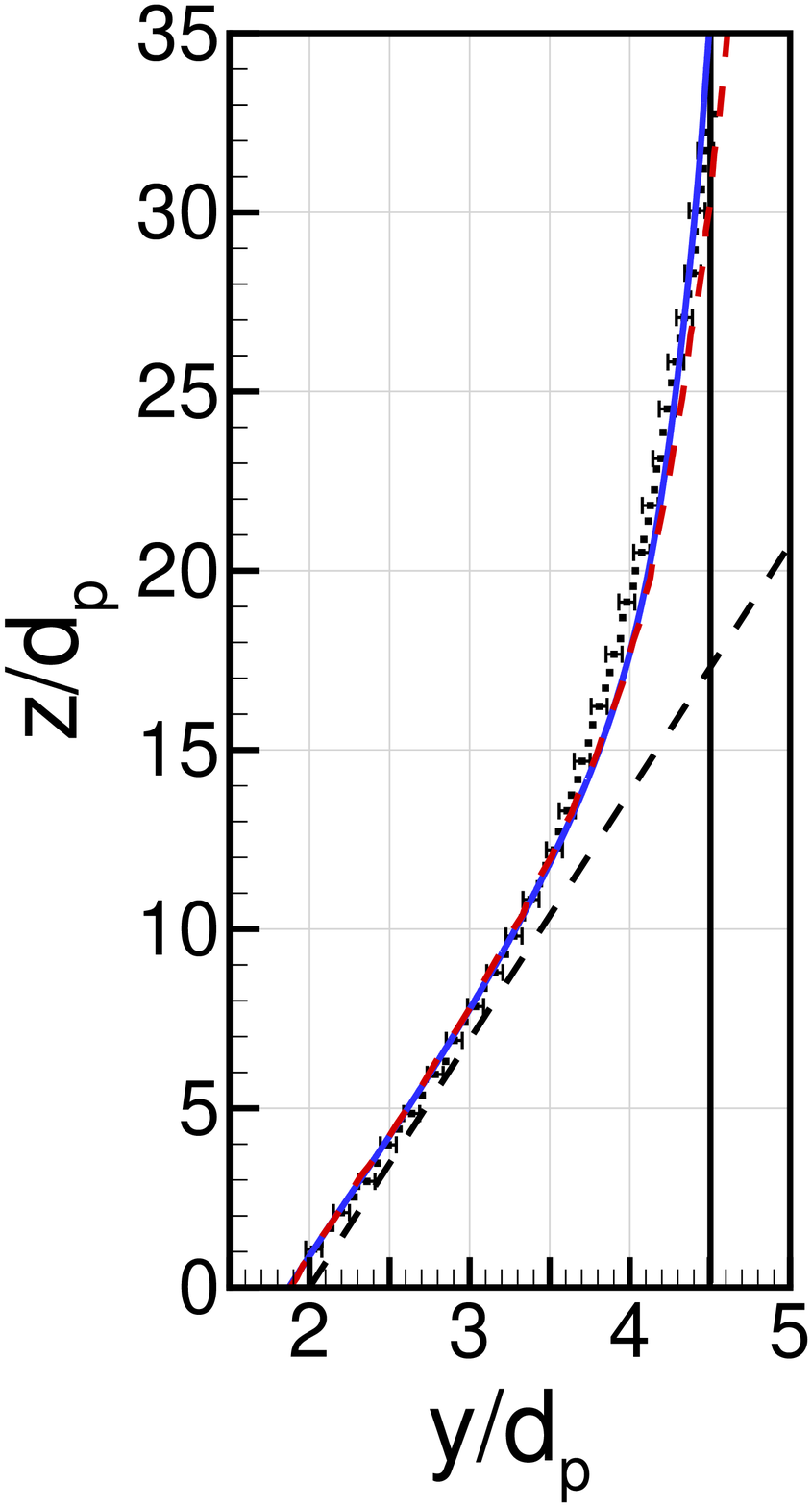} \label{fig:inclined1}}
  \subfigure[]{ \includegraphics[clip, width=37mm]{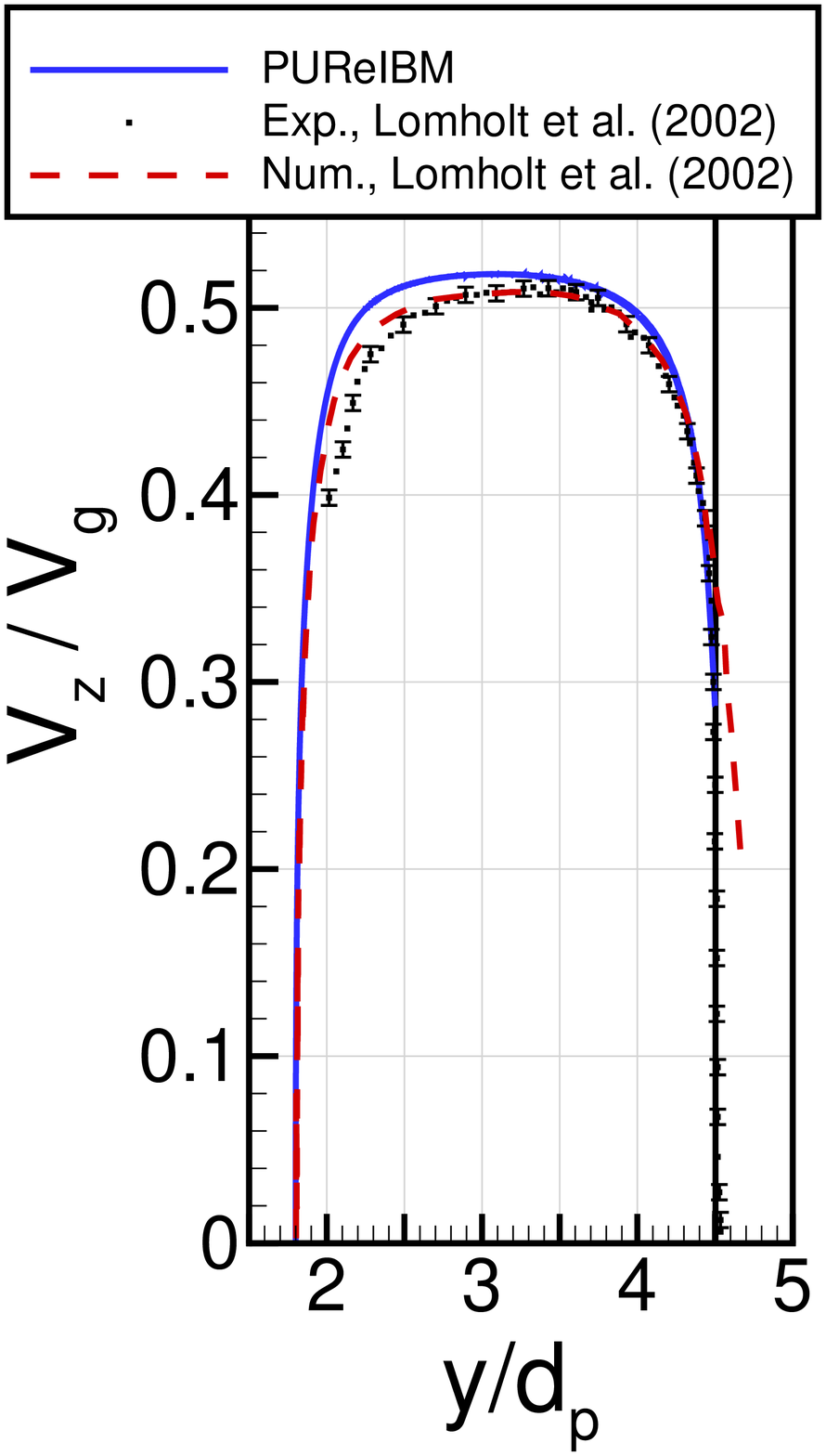} \label{fig:inclined2}}
  \subfigure[]{ \includegraphics[clip, width=37mm]{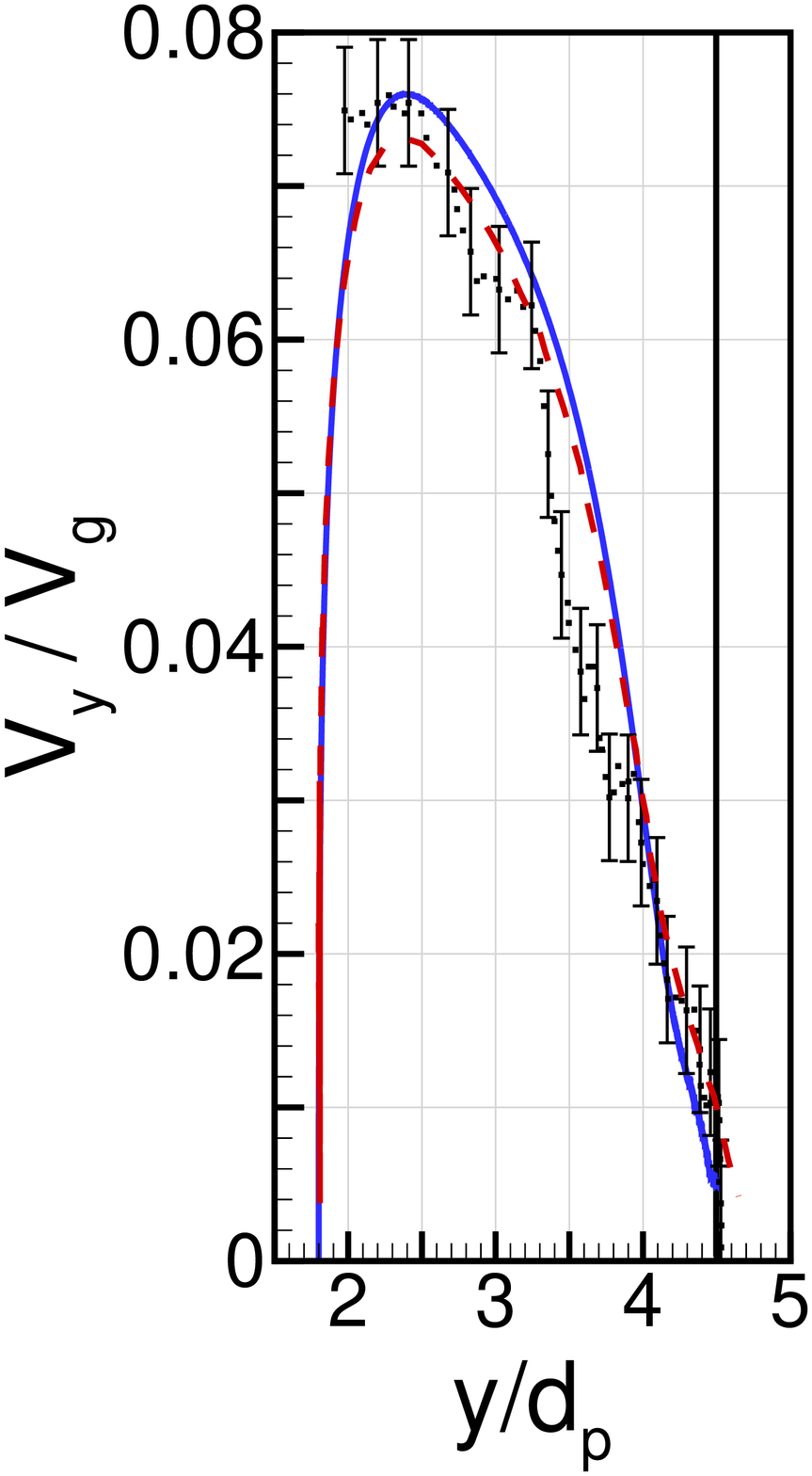} \label{fig:inclined3}}
  \caption{Comparison of predicted results by the present scheme with the experimental study and numerical modeling using force coupling method by \protect\citet{lomholt_2002}.
A solid vertical line at $y/d_p = 4.5$ represents the distance of one particle radius from the wall.
  \subref{fig:inclined1} Particle trajectory (dashed line shows the gravity direction).
  \subref{fig:inclined2} Velocity of the particle in the vertical direction.
  \subref{fig:inclined3} Velocity of the particle in the lateral direction.}
\label{fig:inclined}
\end{centering}
\end{figure}

Fig. \ref{fig:inclined} compares the results from PUReIBM and experimental work and numerical modeling using force coupling method by \cite{lomholt_2002}.

\end{appendix}

\section*{References}

\bibliography{mybibfile}

\end{document}